\documentclass[final,3p,times,twocolumn]{elsarticle}
 \biboptions{comma,sort&compress}
\usepackage{here}
 \usepackage{graphicx}
  \usepackage{epsfig}
\def\nin{\noindent}
\def\beq{\begin{equation}}
\def\eeq{\end{equation}}
\def\bea{\begin{eqnarray}}
\def\eea{\end{eqnarray}}

\journal{Nuc. Phys. (Proc. Suppl.)}

\begin{document}

\begin{frontmatter}
\title{QCD MONOPOLES ON THE LATTICE AND GAUGE INVARIANCE}
 \author[label1]{Adriano Di Giacomo\corref{cor1}}
  \address[label1]{Dipartimento di Fisica Universita' Pisa and INFN Sezione di Pisa\\
3  Largo B. Pontecorvo, 56127 PISA (Italy)}
\cortext[cor1]{Speaker}
\ead{adriano.digiacomo@df.unipi.it}


\begin{abstract}
\noindent
The long standing problem is solved why the number and the location of monopoles
observed  in Lattice configurations depend on the choice of the gauge used to
detect them,  in contrast to the obvious requirement that monopoles, as 
physical objects, must have a gauge-invariant status.
  It is proved, by use of non-abelian Bianchi identities, that monopoles are
indeed gauge-invariant: the technique used to detect them has instead an 
efficiency which depends on the choice of the abelian projection, in a known and 
controllable way.
\end{abstract}

\begin{keyword}
QCD, Confinement, Monopoles
\end{keyword}

\end{frontmatter}

\section{Introduction and motivation}
\nin
Monopoles play a fundamental role in $QCD$: they can condense in the vacuum producing dual superconductivity and confinement\cite{'tHp} \cite{m}.

Much activity has been devoted during the years to observe monopoles in lattice configurations   in the attempt to construct an effective monopole action, the basic idea being that monopoles can be the dominant degrees of freedom of the system. This is expected to be the case at the deconfining transition,
 if the mechanism for confinement is monopole condensation, but it is empirically observed to occur also at lower temperatures, down to zero (Monopole dominance \cite{sco,suz,pol,scol}).
 
 The procedure to detect monopoles in lattice configurations is based on the work of Ref.\cite{dgt} on $U(1)$ gauge theory.
 Any excess over $2\pi$ of the phase of a plaquette indicates that the plaquette is crossed by a Dirac string: a non zero magnetic charge exists in an elementary cube if a net number of Dirac flux lines crosses the plaquettes at its border. In this model the phase of the plaquettes is gauge invariant, and hence the procedure is well defined and gauge invariant.
 In the case of a non abelian gauge theories one has first to fix a gauge and then apply the procedure 
 to an appropriate $U(1)$  subgroup generated by some diagonal component of the Lie algebra\cite{'tH2}. The result depends on the choice of the gauge. The existence of a monopole inside a given elementary cube of the lattice is a gauge dependent feature, and this is of course physically unacceptable.
 
 The prototype monopole configuration in a non abelian gauge theory is the soliton solution of Refs.'s\cite{'tH,Pol}, in the Higgs broken phase of an $SO(3)$ Higgs  model. In that model the $U(1)$ subgroup 
 related to the magnetic charge is the invariance subgroup of the vacuum expectation value of the Higgs field which breaks the symmetry.
 
 In $QCD$ there is apparently no obviously privileged $U(1)$ subgroup. It was proposed in Ref.\cite{'tH2} that any operator in the adjoint representation could be used as an effective Higgs field to define monopoles, the  idea being that  physics is in some way independent of that choice. Each specific choice is called an "Abelian Projection".
 In practice it turned out that different abelian projections have different number and locations of the monopoles. There was a wide consensus on the choice of the so called "Maximal Abelian gauge", as the most convenient to expose abelian dominance and monopole dominance. More recent work \cite{scol} showed that monopole dominance is a common feature of many different abelian projections.

  A relevant step towards the solution of the question was done in  Ref.\cite{bdlp}.  By use of the non abelian Bianchi identities it was shown that the violations of abelian Bianchi identities in any abelian projection are projections on the corresponding fundamental weights of the gauge covariant non-abelian Bianchi identities. We briefly recall the main argument in Section 2.
  
  The physical point is that 
  for each magnetically charged configuration there exists a privileged direction in color space, that of the magnetic monopole term in the multipole expansion at large distances, which is always abelian\cite{coleman} . That direction coincides \cite{bdlp} with
  that of the residual $U(1)$ symmetry in the maximal abelian gauge [Section 3].  Only in that projection, modulo gauge transformations which tend to the identity at large distances, the magnetic charge obeys the Dirac quantization condition, which is at the basis of the detection procedure of Ref.\cite{dgt}. In other gauges the magnetic charge is smaller than the true monopole charge, and monopoles can escape detection. We will show that using as a test ground the soliton solution of Ref.'s\cite{'tH,Pol} [Sect. 3]
  All this is based on the results of Ref.\cite{bdlp} and of a paper in preparation \cite{bdd}.
  In Sect.4 we discuss gauge invariance of dual superconductivity.
 	 In Sect.5 we discuss a numerical test of our statements on $QCD$ lattice configurations \cite{bdd}.
	In Section 6 we summarize our results.
  \section{Non-abelian Bianchi Identities}
 \nin
 In $U(1)$ gauge theory the abelian Bianchi identities  for $F^*_{\mu \nu}\equiv \frac{1}{2} \epsilon_{\mu \nu \rho \sigma} F_{\rho \sigma}$
 \begin{equation}
 \partial_{\mu} F^*_{\mu \nu} =0
  \end{equation}
  are the homogeneous Maxwell's equations:  $\vec \nabla \vec B =0$,
  
    $\vec\nabla\wedge\vec E +\partial_{t} \vec B =0$  and mean non existence of magnetic charge. A violation of the Bianchi identities 
  \begin{equation} 
    \partial_{\mu} F^*_{\mu \nu} = j_{\nu} \label{abi}
    \end{equation}
    corresponds to a non-zero magnetic current $j_{\nu}$, which is conserved because of the antisymmetry of $F^*_{\mu \nu}$.
    
    The non-abelian version is 
    \begin{equation}
    D_{\mu} G^*_{\mu \nu} = J_{\nu} \label{nabi}
    \end{equation}
    with $D_{\mu}$ the covariant derivative and  $G^*_{\mu \nu}$ the dual of the non abelian field strength $G_{\mu \nu}$.  Eq.(\ref{nabi}) implies   $D_{\mu} J_{\mu}=0$.
    
    To extract the full gauge invariant content of Eq.s(\ref{nabi}) one can diagonalize them by a gauge transformation: this is always possible due to the theorem of Ref.\cite{CM} (valid in absence of super-symmetry) stating that the components of the current commute with each other. One can then project on a complete set of independent diagonal matrices, which, by definition of rank, are as many as the rank $r$ of the gauge group. A convenient choice for it are the fundamental weights \cite{bdlp} $\phi^a_0$ $(a= 1,..r)$, one for each simple root of the algebra,  which are defined by their algebraic relations to the commuting operators of the Cartan algebra $H_i$, $(i=1,..r)$, and to the non diagonal operators $E_{\pm \vec \alpha}$ corresponding to the roots $\vec \alpha$ .  
    $[\phi^a_0,H_i] =0$, $[\phi^a_0, E_{\pm \vec \alpha}] \pm (\vec c^a \vec \alpha) E_{\pm \vec \alpha}$,
    $ \vec c^a \vec \alpha^b =\delta_{ab}$ for the simple roots $\vec \alpha^b$, $ (b = 1, ..r)$.
    
    We obtain
    \begin{equation}
    Tr ( \phi^a_{I} D_{\mu} G^*_{\mu \nu}) =  j^a_{\nu}(I) \equiv Tr (\phi^a_{I} J_{\nu} ) \label{de}
    \end{equation}
    
    with $\phi^a_{I}$ the matrix which coincides with the $a-th$ diagonal fundamental weight in the representation in which the currents $J_{\nu}$ are diagonal.
    
    One may also project on $ \phi^a_{V} $  which differs from  $\phi^a_{I}$ in that  $ \phi^a_{V} = V \phi^a_0 V^{\dagger}$ in the representation in which $J_{\mu}$ are diagonal. 
    \begin{equation}
    Tr ( \phi^a_{V} D_{\mu} G^*_{\mu \nu}) =  j^a_{\nu}(V) \equiv Tr (\phi^a_{V} J_{\nu} ) \label{dep}
    \end{equation}
    For $V=I$ one recovers Eq.(\ref{de}).

       $\phi^a_{I}$ is the effective Higgs in the abelian projection in which $J_{\mu}$ are diagonal.  $ \phi^a_{V}$ the effective Higgs in the generic
      abelian projection, in which $J_{\mu}$ is not diagonal.
       We shall denote by  $F^a_{\mu \nu} (V) $ the $a-th$ 'tHooft tensor for that abelian projection.
       
    In Ref.\cite{bdlp} the following theorem is proved
    
    THEOREM : For a generic  compact gauge group and for any arbitrary gauge trnsformation $V$ as a consequence of Eq.(\ref{nabi}) $ \partial_{\mu} {F^a}^*_{\mu \nu}(V) = j^a_{\nu}(V)$.
    
    The abelian magnetic current in a generic abelian projection is the projection on the corresponding effective Higgs of the non abelian magnetic current. 
    
    In what follows we shall  mainly refer to the gauge group $SU(2)$ for the sake of simplicity. Whatever we say, however,  is valid for a generic group. For details see Ref.s\cite{bdlp}\cite{DLP}.
    
    \section{ The soliton monopole}
    The soliton monopole of Ref.s\cite{'tH,Pol} is the prototype configuration in non abelian gauge theories with non zero magnetic charge. It exists in the Higgs-broken phase of the SO(3) Higgs model
    
    $L = -\frac{1}{4}\vec G_{\mu \nu}\vec G_{\mu \nu} + (D_{\mu} \phi)^{\dagger}(D_{\mu} \phi) -V(\phi^2)$
    
    $V(\phi^2)$ is the quartic potential depending on the mass square $\mu^2$, which is negative in the broken phase, and on the quartic-term coupling $\lambda$.
    The original solution was worked out in the so called hedgehog gauge, in which the Higgs field at the position $\vec r$ has in color space the orientation $\hat r \equiv \frac{\vec r}{r}$ .
    
    $ \vec \phi(\vec r) = H(r) \hat r$  , $H(r)_{r \to \infty}\to$ v, the $vev$ of the Higgs field.

    The solution reads 
    
    $\vec A_0 =0$  \hspace{0.5cm}        $ A^a_i= -\epsilon_{iak} \frac{r^k}{gr^2} [ 1 - K(gvr) ] $
    
   $ K(x)$ is a function whose details depend on the specific choice of the parameters in the potential $V(\phi^2)$, but obeys in any case, modulo possible logs, the following conditions
    
    $[ 1 - K(x)]_{x\to 0}\propto x^2 $ ,\hspace{1.5cm}      $K(x)_{x\to \infty} \approx \exp(-x)$
    
    It is trivial to check that in this gauge\cite{bdd}  $\partial_{\mu} A_{\mu} =0$ : the hedgehog gauge is nothing but the Landau gauge!
    
    The abelian magnetic field in this gauge can be computed  as $b_{i} = \frac{1}{2} \epsilon_{i j k }(\partial_{i}A^3_{j} - \partial_{j} A^3_{i}) $  giving  $\vec b \approx_{r \to \infty}  \frac{2\hat r}{gr^2}cos(\theta)$.  The magnetic charge is the flux of $\vec b$ through the surface at infinity
    and is trivially zero.  
    
    One can gauge-transform to the unitary gauge,  where the Higgs field $\vec \phi$ is directed along the 3-axis in color space. The gauge transformation is singular at $\vec r= 0$, but can
    be regularized \cite{shnir} and the field  can be explicitly computed.
    It can then be checked \cite{bdlp} by explicit calculation that this gauge is nothing but the maximal abelian gauge, defined by the condition  
    \begin{equation}
     \partial_{\mu}A_{\mu}^{\pm} \pm ig \left[A_{\mu}^3, A_{\mu}^{\pm} \right] =0 \label{mag}
     \end{equation}
     The solution is static, so that $J_{i} =0$ , and only $J_{0}$ is non-zero. The identities Eq.(\ref{nabi})
     read
    \begin{equation}
J_{0} = D_{i}B_{i} = \frac{2\pi}{g}\delta^3(\vec r)\sigma_3
\end{equation}
and projecting on the fundamental weight  $\phi_{0}= \frac{\sigma_3}{2}$ our theorem gives the abelian Bianchi identity 
\begin{equation}
\vec \nabla \cdot \vec b = \frac{2\pi}{g} \delta^3(\vec r)
\end{equation}
or, for the magnetic charge
\begin{equation}
Q_m = \frac{1}{g}
\end{equation}
Not all of the abelian projections are equivalent! The same monopole has zero charge in the Landau gauge, charge 2 in the maximal abelian gauge, where it obeys the Dirac quantization condition.

 This is because the configuration has a privileged direction in color space, that of the magnetic field strength at large distances.
 
  The argument can be extended to  generic static configurations, by  use of a theorem of Coleman\cite{coleman}: 
   The magnetic monopole term in the multipole expansion of a generic static field configuration is abelian: it obeys abelian Eqs. of motion and can be gauged along one direction in color space (modulo a global transformation).

 For a generic configuration the theorem holds for the superposition of the state with its time reflected, with the same consequences.

 The magnetic field at large $r$ in this gauge only depends on the total magnetic charge. 
       $\vec B = \frac{m}{2} \frac{\vec r}{2gr^3}\sigma_3 $  .  
       
       Terms in $\sigma_{\pm} $ are non leading in $r$ at large $r$, since they belong to higher multipoles.  As a consequence the abelian magnetic field corresponding to the 'tHooft tensor is
        $\vec b \approx_{r\to \infty}  \frac{m}{2}\frac{\vec r}{2gr^3}$
        and the total magnetic charge $\frac{m}{2g}$.
    
  The gauge field at large distances obeys the  gauge condition Eq.(\ref{mag}). This gauge can differ from the max abelian by a gauge  transformation $W(\vec r)$  with $W(\vec r )_{r \to \infty}\to I$.     

Starting from the maximal abelian gauge one can perform a gauge transformation depending on one parameter ${\bf a}$, of the form:
\begin{equation}
U_{\bf a}(\theta, \phi) = \exp(i\phi\frac{\sigma_{3}}{2}) \exp(i\theta {\bf a} \frac{\sigma_{2}}{2}) \exp(-i\phi\frac{\sigma_{3}}{2}) \label{gt}
\end{equation}
For ${\bf a} =0$  $U_{0}(\theta, \phi)=1$ and one stays in the maximal abelian gauge, for ${\bf a} =1$ $U_{1}(\theta, \phi)$ transforms from Maximal  Abelian  to Landau gauge \cite{shnir}.   
The expected magnetic charge is 
\begin{equation}
\frac{Q_{m}({\bf a})}{Q_{m}(0)} =\frac{1 + \cos({\bf a} \pi)}{2} \le 1 \label{gtm}
\end{equation}
This prediction is being checked on the lattice \cite{bdd}.

\section{Monopole condensation and confinement}
\nin
 The magnetic charge density, in the maximal-abelian  projection  is
\begin{equation}
                       j_0(x,I) = Tr(\phi_{I} J_0(x))
  \end{equation}                     

 with $J_0$  defined in Eq.(\ref{nabi}) the non abelian magnetic current, and
 $\phi_{I}$ the fundamental weight  diagonal with it.

 The equal-time commutator   with any local operator $O(y)$ carrying magnetic charge $m$ is
  
\begin{equation}
 \left[ j_0(\vec x, x_0,I), O(\vec y, x_0)\right] = m \delta^3(\vec x -\vec y)O(\vec y,x_0)  + S.T.
  \end{equation}
  
 or, after integration over $\vec x$ 
\begin{equation}
 \left[Q(I),O(y)\right] = m O(y) 
\end{equation}
 If $m \neq 0$ and $\langle O \rangle \neq 0$ the magnetic $U(1)$ is Higgs broken (dual superconductivity).

  Consider now the magnetic charge density in a generic abelian projection 
  \begin{equation}
   j_0(x,V) = Tr(V(x)\phi_{I} V^{\dagger}(x)J_0(x))
   \end{equation}
    and
 compute the trace in the gauge in which $J_0(x)$ is diagonal and $\phi_{I}$ with it.
 
 Since $ \phi_{I} $ and $V(x) \phi_{I} V^{\dagger}(x)$ belong to the algebra it will be
 \begin{equation}
 V(x) \phi_{I} V^{\dagger}(x) = C(x,V) \phi_{I} +\sum  _{\vec \alpha} E_{\vec \alpha} D^{\vec \alpha}(x,V) 
 \end{equation}
 
 In the gauge chosen only the first term will contribute to the trace and, as a consequence,
 \begin{equation}
  \left[Q(V),O(y)\right] = m O(y)C(y,V)
  \end{equation}
  Since $C(y,V) $ is generically non vanishing , $O$ will have a non-zero charge also in the new gauge 
  and if  $\langle O \rangle \neq 0$, also the new $U(1)$ is Higgs broken.
    A consequence of the non-abelian Bianchi identities Eq.(\ref{nabi}) is that the system behaves as a superconductor in all abelian projections.  Dual superconductivity is a gauge-invariant property.
    \section{Monopoles on the lattice}
    \nin 

  On the lattice monopoles are assumed to be point-like and to lie inside the elementary cubes . Their number density being finite, the probability of finding  them on  the border of the cube is zero. Looking at the flux through cubes of different sizes can help to eliminate ultraviolet noise \cite{ddmo}. 
  
  In the Maximal-abelian gauge monopoles are detected by the method of Ref.\cite{dgt} : the surface of the cube on which the flux is measured is assumed to be the sphere at spatial infinity.
 Any gauge transformation which is trivial on the sphere at infinity (on the border of the cube) is irrelevant. 
 A gauge transformation $U(\theta, \phi)$ on the sphere can change the number of observed monopoles.  
 As an example the gauge transformation Eq.(\ref{gt}) can change the observed charge as predicted by Eq.(\ref{gtm}).  On can assume that the monopole is, on the average, at the centre of the cube and take as polar axis the direction normal to the plaquette which is crossed by the Dirac string. This introduces discretization errors, but the qualitative behavior is expected to be that of Eq.(\ref{gtm}).  A  Lattice study on the way\cite{bdd}.
    \section{Concluding remarks}
    We can summarize our results as follows:
    
    1) It is not true that any operator in the adjoint representation can act as an effective Higgs field to define the abelian projection and  monopole charge. Each magnetically charged configuration has its own preferred direction in color space, which is fixed by the magnetic monopole term in the large distance multipole expansion of the field. This term can be gauged to a fixed direction in color space.
    
    2) This direction is selected by the Maximal-Abelian gauge and identifies the residual gauge symmetry.
     As a consequence magnetic charge can be measured, e.g. on the lattice, by use of the technique of
     Ref.\cite{dgt} in this gauge, and defines gauge invariant monopoles. If one tries to detect these monopoles by the same technique in other abelian projections, one usually finds a smaller charge.  This is not because the monopole is not there,
     but only because one is looking at something which is not the magnetic charge, which is only defined by the multipole expansion.
     
     3) In particular the magnetic charge in the Landau gauge is zero. We find that this is indeed the case on the lattice in agreement with previous observations \cite{suzu}.
     
     4) Dual superconductivity is a gauge invariant feature. If any charged field produces condensation 
     by acquiring a non-zero $vev$ it will be magnetically charged with respect to all $U(1)$'s defined by any abelian projection.
     
     5) What we have presented refers to $SU(2)$ , but it applies to any compact gauge group. Actually all the results were obtained for the generic case in Ref.s\cite{DLP}\cite{bdlp}\cite{bdd}.

\end{document}